
\input phyzzx
\input tables
\overfullrule 0pt
\tolerance 1600
\hbadness 1600
\vbadness 1600
\linepenalty=100
\REF\i {S.~Ritter and J.~Ranft, Acta Phys. Polonica B11 (1980)
259; S.~Ritter, Z. Phys. C6 (1982) 27;
Comp. Phys. Comm. 31 (1984) 393.}
\REF\ii {K. H\"an\ss gen and S.~Ritter, Comp. Phys. Comm.
31 (1984) 411.}
\REF\sj {H.-U.~Bengtsson and T.~Sj{\"o}rstrand, Comp. Phys. Comm.
46 (1987) 43.}

%
\REF\iii {J. Ranft, Proceedings Multiparticle Dynamics 1988, D. Schiff and
J. Tran Thanh Van (ed.), Editions Frontiers, 1988, p. 331. }
 \REF\iiii {J.~Ranft, P.~Aurenche, F.~Bopp,
A.~Capella, K.~Hahn, J.~Kwiecinski, M.~Maire and J.~Tran Thanh Van,
 SSC report SSC-149 (1987).}
\REF\iiiii  {K. Hahn and J. Ranft, Phys. Rev. D41 (1990)  1463. }
\REF\Aurenche{
P.~Aurenche, F.~W. Bopp, A.~Capella, J.~Kwiecinski, M.~Maire, J.~Ranft  and
  J.~Tran Thanh~Van,
Phys.~Rev.~D45 (1992) 92.}

\REF \Bopp{
F.~W. Bopp, A.~Capella, J.~Ranft  and J.~Tran Thanh~Van,
Z.~Phys.~C51 (1991) 99.}

\REF\Pertermann{
F.~W. Bopp, D.~Pertermann  and J.~Ranft,
Z.~Phys.~C54 (1992) 683.}

\REF\Engel{
R.~Engel, F.~W. Bopp, D.~Pertermann  and J.~Ranft,
Phys.~Rev.~D46 (1992) 5192.}

\REF\Roesler{
S.~Roesler, R.~Engel  and J.~Ranft,
Z.Phys.C159 (1993) 481.}

\REF\BPER{
F.~W. Bopp, D.~Pertermann, R.~Engel and J.~Ranft, INFN-preprint (1993)
submitted to  Phys. Rev. D .}

\REF\RBEP{
J.~Ranft, F.~W. Bopp, R.~Engel, D.~Pertermann and S.~Roesler
to be published in the proceedings of the International Conference
on Monte-Carlo-Simulation in High Energy and Nuclear Physics,
Tallahassee FL (1993).}

\REF\RCT{
J.~Ranft, A.~Capella, J.~Tran~Thanh~Van, Lund preprint Lu-TP 93-18, to appear
in Phys.Lett.B (1993).}

\REF\Capella{ A review is given in:
A.~Capella, U.~Sukhatme, C.~I. Tan  and J.~Tran Thanh~Van,
Phys. Rep. 236, no.4,5 (1994) 227. }

\REF\v {A.~Capella and J.~Tran Thanh Van, Z. Phys. C10 (1981) 249;
Phys. Lett. 114B (1982) 450.}
\REF\vi {P.~Aurenche and F.W.~Bopp, Z. Phys. C13 (1982) 205;
Phys. Lett. 114B (1982) 363.}
\REF\vii {A.B.~Kaidalov, Phys. Lett. 116B (1982) 459;
A.B.~Kaidalov and K.A.~Ter-Martirosyan, Phys. Lett. 117B (1982) 247.}
\REF\viii {J.~Ranft, P.~Aurenche, and F.W.~Bopp,
Z. Phys. C26 (1984) 279;
P.~Aurenche, F.W.~Bopp, and J.~Ranft, Z. Phys. C23 (1984) 67;
P.~Aurenche, F.W.~Bopp, and J.~Ranft, Phys. Lett. 147B (1984)
212.}
\REF\ix {P. Aurenche, F.W. Bopp, and J.~Ranft,
Phys. Rev. D33 (1986) 1867.}
\REF\x {G. Arnison et al., UA1 Collaboration,
Phys. Lett. 118B (1982) 167.}
\REF\xii {P.~Aurenche, F.W.~Bopp, and J.~Ranft,
Contribution to the Workshop on Physics Simulations
at High Energy, Madison, Wisconsin, 1986, Annecy preprint
LAPP-TH-161 (1986).}
\REF\xiii {A.~Capella, J.~Tran Thanh Van, and J.~Kwiecinski,
Phys. Rev. Lett. 58 (1987) 2015.}
\REF\xiv {L.~Durand and H. Pi, Phys. Rev. Lett. 58 (1987) 303.}
\REF\xv {A.D. Martin, R.G. Roberts and W.J. Stirling, Phys. Rev. D 37 (1988)
116.}

\REF\KMRS{
J.~Kwiecinski, A.~D. Martin, R.~G. Roberts  and W.~J. Stirling,
Phys.~Rev.~D42 (1990) 3645.}

\REF\MRSII{
A.~D. Martin, R.~G. Roberts  and W.~J. Stirling,
Phys.~Rev.~D47 (1993) 867.}

\REF\CTEQ{J. Botts et al. (CTEQ-collaboration), Phys. Lett. 304B (1993) 159.}

\REF\BEKPR{J.~Ranft, F.~W. Bopp, R.~Engel, I.~Kawrakow and  D.~Pertermann,
INFN-prepint (1993), presented at the LBL workshop on preequilibrium
parton dynamics (Sept. 1993) Berkeley. }



\rightline{SI-93-3}
\rightline{Jan. 1994}

\vskip 0.75in
\centerline{\bf DTUJET--93}
\vskip .5cm

\baselineskip = 14pt
\centerline{\bf Sampling inelastic proton--proton and antiproton--proton}
\centerline{\bf collisions according to the two--component Dual Parton Model}

\vskip 0.5in
\centerline{
P.~Aurenche \foot{LAPP, F--74941 Annecy-le-vieux, France}),
F.~W.~Bopp
\foot{Fachbereich Physik, Universit\"at Siegen, D--57068 Siegen, Germany}),
}
\centerline{
R.~Engel
\foot{Fachbereich Physik, Universit\"at Leipzig, D--04109 Leipzig, Germany}),
D.~Pertermann$^{\dag }$),
J.~Ranft
\foot{INFN,
Lab. Naz. di Frascati, Via Enrico Fermi 40, I-00044 Frascati, Italy}),
and
S.~Roesler$^{\dag }$)}
\vskip 1in

\centerline{\bf Abstract}

A new version of a Monte Carlo Program for hadronic multi-particle production
is presented. It is based on the two-component Dual Parton Model which includes
the dual topological unitarization of soft and hard cross sections. The model
treats both soft (low $p_{\perp}$) and hard
(minijet, large $p_{\perp}$) processes in a unified
and consistent way.  The unified description is important
at  TeV-energies of hadron colliders, where the
hard perturbative cross sections of QCD become large
and comparable to the total cross sections.

\vfill\eject
\chapter{Program Summary}

$$\vbox{
        \halign{#\hfil\quad &
                \vtop{\hsize 3.5in \strut # \strut}\cr
Title of program:               & DTUJET \cr
\noalign{\vskip 10pt}
Catalog number:                 & \cr
\noalign{\vskip 10pt}
Program obtainable              & from CPC Program Library,\cr
                                & Queen's University of Belfast, N. Ireland\cr
                                & (see application form in this issue)\cr
\noalign{\vskip 10pt}
Computer on which the program   & RISC--Workstations: \cr
has been thoroughly tested:     & DEC, HP, and IBM  \cr
\noalign{\vskip 10pt}
Operating system:               & UNIX--systems:      \cr
                                & DEC--ULTRIX, HP--UX and IBM--AIX\cr
\noalign{\vskip 10pt}
Programming language used:      & FORTRAN-77\cr
\noalign{\vskip 10pt}
Memory required to execute      & 11.5 megabyte \cr
with typical data using the     & \cr
{\tt DOUBLE PRECISION} options  & \cr
\noalign{\vskip 10pt}
No. of bits in a word           & 64 \cr
 \noalign{\vskip 10pt}
Has the code been vectorized?   & no \cr
\noalign{\vskip 10pt}
Number of lines                 & 45000 (including data files)\cr
in distributed program:         & \cr
\noalign{\vskip 10pt}
Other programs used in DTUJET  & BAMJET \refmark\i \cr
in modified form               & Sampling the hadronization of strings.\cr
                                & DECAY \refmark\ii \cr
                                & Sampling the decay of hadron resonances.\cr
                                & JETSET \refmark\sj \cr
                                & Sampling the decay of strings.\cr
                                & RNDM  \cr
                                & Processor independent random number\cr
                                & generator.\cr
}   
}$$ 

\vfill\eject
$$\vbox{
        \halign{#\hfil\quad &
                \vtop{\hsize 3.5in \strut # \strut}\cr

Keywords:                     & Monte Carlo event generator,\cr
                              & hadron-hadron scattering, \cr
                              & fragmentation, hadronic final states.\cr
\noalign{\vskip 10pt}
Nature of physical problem:   & Simulation of high energy hadron collisions.\cr
\noalign{\vskip 10pt}
Method of solution:           & Monte Carlo generation of individual events\cr
                              & which step by step follows the evolution \cr
                              & of the scattering process postulated \cr
                              & in the Dual Parton Model. \cr
\noalign{\vskip 10pt}
Restrictions on the complexity & Above 40~TeV arrays have to be enlarged.\cr
of the problem                & No tests were done below $\sqrt s = 50~$GeV.\cr
\noalign{\vskip 10pt}
Typical running time:         & 100 thousand events at an energy of 40 TeV \cr
                              & require a DEC--station 5260 running time \cr
                              & of 33 minutes, of which  about 30 seconds \cr
                              & are used for initialization. \cr
}   
}$$ 

\vfill\eject

\chapter{Dual Unitarization of Soft and Hard Hadronic Interactions}

The dual topological unitarization of hard and soft hadronic cross
sections is the basis for our model of hadronic multi-particle production.
First descriptions of the model were given in [{\iii - \iiiii}].
The physics of the former version of the code, DTUJET--90, which is
largely unchanged,  was described in the papers [{\Aurenche,\Bopp}].
The program follows the evolution of the scattering process step by step
in a probabilistic way. It maintains energy, momentum and flavor conservation
with an relative accuracy of better than $10^{-4}$~.

New in the code DTUJET--93 is the use of modern parton structure functions
for calculating the minijet component, new fits to total, elastic, inelastic,
and single diffractive cross sections, a careful treatment of the low--mass
and high--mass single diffractive component
\refmark{\Pertermann,\Engel,\Roesler} and the introduction of sea diquarks
and other improvements in order to get a updated description of strange
particle production in the Dual Parton Model \refmark\RCT \ .

Soft hadronic multi-particle production in the framework
of the Dual Parton Model \refmark{\Capella}  was studied  by several
groups \refmark{\v-\ix}.
These studies and in particular the Monte-Carlo-formulation
of the model in the form of a dual
multi-chain fragmentation model \refmark{\viii,\ix}\ were the
first starting point of DTUJET model.

Experimental observations at collider energies made
it clear that the soft and hard components of hadronic
multi-particle production are closely related.  We refer
to the discovery of correlations between the average
transverse momentum of produced hadrons and the multiplicity
density in the central rapidity range \refmark\x .
This property could be understood within the dual
multi-chain fragmentation model \refmark{\ix,\xii}\ only by introducing
transverse momenta at the ends of the fragmenting strings.
The magnitudes of some of the transverse momenta needed in this way
could only be interpreted as due to hard constituent scattering.

The need for a uniform treatment of hard and soft hadronic
multi-particle production is furthermore underlined by the
fact that the perturbative QCD cross section for hard
constituent scattering rises strongly with the energy. It reaches
values  larger
than the total hadronic cross section at the energies of proposed
supercolliders.
At these energies one expects
that unitarity corrections should play an important role.
Those corrections then inevitably lead to several semi-hard
interactions resulting in an increase of the average number of
jets.

The perturbative hard constituent scattering is  one of the
processes responsible for the rise of the hadronic cross sections.
This was studied quantitatively in papers by Capella,
Tran Thanh Van and Kwiecinski \refmark\xiii\ and by
Durand and Pi \refmark\xiv, where the consequences
for the total and inelastic cross sections of the unitarization
of soft and hard scattering cross sections were studied.  This
model in the form as formulated in [\xiii ]\ is the
second starting point of the DTUJET-model.

For more details about the model we refer to the papers
[{\iiii-\RCT}]\ .

\bigskip

\chapter{Description of ~DTUJET--93}
\section{The Structure of DTUJET--93}

The DTUJET source code consists of several FORTRAN files.
The user will need the following ones:

\hangafter = 1
\hangindent = 1.5in
\parindent = 0pt
\leavevmode
\hbox to \hangindent
{\hskip 25pt $\bullet$ \tt dtu93mai:\hfil}contains the routines
which organize the various tasks to be performed. The central part,
the main program DTMAIN, will be described in the next section.

\hangafter = 1
\hangindent = 1.5in
\parindent = 0pt
\leavevmode
\hbox to \hangindent
{\hskip 25pt $\bullet$ \tt dtu93col:\hfil}collects actual events
utilizing the following files.
Contains call to default and dummy histograming routines.

\hangafter = 1
\hangindent = 1.5in
\parindent = 0pt
\leavevmode
\hbox to \hangindent
{\hskip 25pt $\bullet$ \tt dtu93pom:\hfil}contains
routines to calculate the multi-Pomeron distributions,
sample from these distributions, and calculate the total and
inelastic cross sections.

\hangafter = 1
\hangindent = 1.5in
\parindent = 0pt
\leavevmode
\hbox to \hangindent
{\hskip 25pt $\bullet$ \tt dtu93par:\hfil}formulates the model on the
parton level, samples the
strings and the partons at the ends of the strings.

\hangafter = 1
\hangindent = 1.5in
\parindent = 0pt
\leavevmode
\hbox to \hangindent
{\hskip 25pt $\bullet$ \tt dtu93diq:\hfil}contains the routines for sampling
sea--sea chains with sea-diquarks besides sea-quarks at the ends.

\hangafter = 1
\hangindent = 1.5in
\parindent = 0pt
\leavevmode
\hbox to \hangindent
{\hskip 25pt $\bullet$ \tt dtu93sof:\hfil}contains the routines for sampling
the soft parton distributions by an alternative method.

\hangafter = 1
\hangindent = 1.5in
\parindent = 0pt
\leavevmode
\hbox to \hangindent
{\hskip 25pt $\bullet$ \tt dtu93lap:\hfil}contains the routines used to
sample the hard scattering of constituents.

\hangafter = 1
\hangindent = 1.5in
\parindent = 0pt
\leavevmode
\hbox to \hangindent
{\hskip 25pt $\bullet$ \tt dtu93tcb:\hfil}contains the BAMJET
\refmark\i\ and DECAY \refmark\ii\
routines for string fragmentation and routines to
test and call BAMJET.

\hangafter = 1
\hangindent = 1.5in
\parindent = 0pt
\leavevmode
\hbox to \hangindent
{\hskip 25pt $\bullet$ \tt dtu93lun:\hfil}contains a short interface
to the JETSET\refmark\sj\ program.

\hangafter = 1
\hangindent = 1.5in
\parindent = 0pt
\leavevmode
\hbox to \hangindent
{\hskip 25pt $\bullet$ \tt dtu93luj:\hfil}A copy of JETSET version 7.3
(converted to allow for compilation under {\tt DOUBLE PRECISION}) is
included for the convenience of the user.

\hangafter = 1
\hangindent = 1.5in
\parindent = 0pt
\leavevmode
\hbox to \hangindent
{\hskip 25pt $\bullet$ \tt dtu93dif:\hfil}contains the routines for sampling
diffractive chains.

\hangafter = 1
\hangindent = 1.5in
\parindent = 0pt
\leavevmode
\hbox to \hangindent
{\hskip 25pt $\bullet$ \tt dtu93his:\hfil}contains standard histograming
routines.

\hangafter = 1
\hangindent = 1.5in
\parindent = 0pt
\leavevmode
\hbox to \hangindent
{\hskip 25pt $\bullet$ \tt dtu93lib:\hfil}contains general
purpose library routines.

\vfill\eject
\parindent=25pt
\noindent
On the UNIX-machines these files have the extension {\tt .f}.
In addition, data files {\tt struf90.dat} and {\tt struf93.dat} will
be required by a
routine in {\tt dtu93lap}.
These files contain tables which allow to calculate structure functions
parametrized in various ways.

Most parts of the program, for instance the

\item{--} sampling of multi-Pomeron events,
\item{--} hadron structure functions,
\item{--} hard perturbative constituent scattering cross sections,
\item{--} the method of string fragmentation into hadrons, and
\item{--} the method of selecting the exclusive parton events
($x$-values and flavors),

\noindent
are programmed as distinct entities, which can in principle be exchanged
by suitable other codes.
For some of these tasks DTUJET--93
offers several options; for others this is foreseen
and easy to implement.

\vfill\eject

\section{The Main Program  }

\leavevmode
The main program DTMAIN is provided to allow for a stand-alone running
of the program. It is unnecessary for a user who wants to call the
event generator from its own programs.

It contains the following lines
\bigskip
\line{\tt \hskip 1.5in     PROGRAM DTMAIN      \hfil}
\line{\tt \hskip 1.3in 1 \  CALL DTPREP(NEVNT) \hfil}
\line{\tt \hskip 1.5in     DO 9 I=1,NEVNT      \hfil}
\line{\tt \hskip 1.3in 9 \ CALL DTCOLL         \hfil}
\line{\tt \hskip 1.5in     CALL DTHIST         \hfil}
\line{\tt \hskip 1.5in     GOTO 1              \hfil}
\line{\tt \hskip 1.5in     END                 \hfil}

\noindent
which are largely self explanatory.

To prepare the generator a call
to {\tt DTPREP} is necessary.
It initializes variables in a way chosen by the user
and allows for preparatory tasks. Its actions are controlled by input
cards. Each input card is identified by a code word which specifies its
function as described in the next section. Without specification default
values are taken. The subroutine is left if the
code word START is encountered.
Please note, that the nature of these initializations makes it for
many tasks impossible to create events,
which need a different initialization during one run
of the code. A good praxis is therefore,
to have a separate run of the code for each problem to be solved.

A call to the subroutine {\tt DTCOLL} generates one
event. The event is stored in
a {\tt COMMON} block {\tt /USER/} and {\tt /PARTCL/} as described in section 4.

To check the generated events, a call to the subroutine {\tt DTHIST} provides
short histogram output, which was prepared in {\tt DTPREP} and {\tt DTCOLL}
without action of the user.
It contains a call to a dummy routine {\tt DTHSTO(I)},
which can be replaced by a histograming routine supplied by the user.
It is called at the beginning with the argument $I=1$ for initialization
and preparatory work, with the argument $I=2$ each time a new event has been
generated to allow summations needed for the calculation of expectation values
and with the argument $I=3$ after all the desired events have been generated
for normalizations and print outs.

\vfill
\bigskip
\section{Input Cards and Options}

\leavevmode
The subroutine DTPREP uses input cards.
All input cards of DTUJET have the following form:

\medskip
\line{\tt \hskip 1.5in CODEWD, (WHAT(I), I = 1,6), SDUM\hfil}
\line{\tt \hskip 1.5in FORMAT ( A8, 2X, 6E 10.0, A8)\hfil}

\noindent
The initial code word, which is
read in as {\tt CODEWD}, determines the meaning of the remaining
variables, as described in the following.

\bigskip
\noindent
Code word:  TITLE

\line{Parameter used:  none\hfil }

\noindent
This card must be followed by a card giving the title of the
run, which will be reproduced in the output.

\bigskip
\noindent
Code word:  COMMENT

\line{Parameter used: \tt NCOM = WHAT(1)\hfil }

\noindent
This card allows one to add {\tt NCOM} comment lines. It must be
followed by {\tt NCOM} comment cards, which will be reproduced
in the output.

\line{The default is \tt NCOM~=~1.\hfil }

\bigskip
\noindent
Code word:  CMENERGY

\line{Parameter used: \tt  ECM~=~WHAT(1)\hfil }

\noindent
This card defines the center of mass energy {\tt ECM} of the collision in GeV.

\line{The default is {\tt ECM~=~1800~}GeV.\hfil }

\bigskip
\noindent
Code word:  PROJPAR

\line{Parameter used: \tt  PROJTY~=~SDUM \hfil }

\noindent
Defines the projectile particle type, defined to move into
positive $z$-direction.

\line{The possible values are: \tt  PROTON, APROTON\hfil }
\line{The default is \tt PROTON\hfil }

\bigskip
\noindent
Code word:  TARPAR

\line{Parameter used: \tt  TARGTY~=~SDUM\hfil }

\noindent
Defines the target particle type, defined to move into
negative $z$-direction.

\line{The possible values are: \tt  PROTON 
\hfil}
\line{The default is \tt PROTON\hfil }

\vfill\eject
\bigskip
\noindent
Code word:  SINGDIFF

\line{Parameters used: \hfil }
\line{~~~~~~~~~~~\tt  ISINGD~=~WHAT(1), IDUBLD~=~WHAT(2), SDFRAC~=~WHAT(3)
\hfil }

\noindent
Calls or suppresses  diffractive events.

\medskip
\line{\ \ \tt  ISINGD \hfill  }

\hangafter = 1
\hangindent = 2.1in
\parindent = 0pt
\leavevmode
\hbox to \hangindent
{\hskip 1.1in =~0~:\hfil}Single diffraction suppressed.

\hangafter = 1
\hangindent = 2.1in
\parindent = 0pt
\leavevmode
\hbox to \hangindent
{\hskip 1.1in =~1~:\hfil}Single diffraction included
to a fraction given by third parameter.

%

\medskip
\line{\ \ \tt  IDUBLD  \hfill  }

\hangafter = 1
\hangindent = 2.1in
\parindent = 0pt
\leavevmode
\hbox to \hangindent
{\hskip 1.1in =~0~:\hfil}Double diffraction included.

\hangafter = 1
\hangindent = 2.1in
\parindent = 0pt
\leavevmode
\hbox to \hangindent
{\hskip 1.1in =~1~:\hfil}Only double diffractive events.

\medskip
\line{\ \ \tt  SDFRAC \hfill  }

\hangafter = 1
\hangindent = 2.1in
\parindent = 0pt
\leavevmode
\hbox to \hangindent
{\hskip 1.1in =~0~..~1~:\hfil}Fraction of single diffractive events
to be included in inelastic events. The value $1$ ($0.05$) means
that all ($5 \%$) of the single diffractive events are included.

\noindent
\line{The defaults are \tt  ISINGD~=~0, IDUBLD~=~0, SDFRAC~=~0~. \hfil }

\vfill\eject

\medskip
\bigskip
\noindent
Code word:  STRUCFUN

\line{Parameter used: \tt  ISTRUF~=~WHAT(1)  \hfil  }

\noindent
Defines the structure functions used in the sampling of hard constituent
scattering  \refmark{\xv, \KMRS, \MRSII, \CTEQ}.

\smallskip
\line{\ \  \tt ISTRUF  \hfil  }

\vskip -0.25cm
\line{\hskip 1.1in = ~3~: \hskip 0.4in Martin, Roberts, Stirling:
Set 1\hfil}

\line{\hskip 1.1in = ~9~: \hskip 0.4in Kwiecinski,Martin, Roberts, Stirling:
Set B0\hfil}

\line{\hskip 1.1in =~10~: \hskip 0.4in Kwiecinski,Martin, Roberts, Stirling:
Set B$_-$\hfil}

\line{\hskip 1.1in =~11~: \hskip 0.4in Kwiecinski,Martin, Roberts, Stirling:
Set B$_{-2}$\hfil}

\line{\hskip 1.1in =~12~: \hskip 0.4in Kwiecinski,Martin, Roberts, Stirling:
Set B$_{-5}$\hfil}

\line{\hskip 1.1in =~13~: \hskip 0.4in Martin, Roberts, Stirling (1992):
Set S0\hfil}

\line{\hskip 1.1in =~14~: \hskip 0.4in Martin, Roberts, Stirling (1992):
Set D0\hfil}

\line{\hskip 1.1in =~15~: \hskip 0.4in Martin, Roberts, Stirling (1992):
Set D$_-$\hfil}

\line{\hskip 1.1in =~16~: \hskip 0.4in CTEQ Collaboration (1993):
Set 1M  \hfil}

\line{\hskip 1.1in        =~17~: \hskip 0.4in CTEQ Collaboration (1993):
Set 1MS \hfil}

\line{\hskip 1.1in        =~18~: \hskip 0.4in CTEQ Collaboration (1993):
Set 1ML \hfil}

\line{\hskip 1.1in        =~19~: \hskip 0.4in CTEQ Collaboration (1993):
Set 1D \hfil}

\line{\hskip 1.1in        =~20~: \hskip 0.4in CTEQ Collaboration (1993):
Set 1L \hfil}

\line{\hskip 0.55in        =~210~..~220~: \hskip 0.4in as above with energy
dependent $p_{\perp}$ threshold value\hfil}

\smallskip
\line{The default is 215 . \hfil }
The fits include functions
with conventional $1/x$ singularity of sea-quarks and
gluon distributions (for instance the MRS[D0] functions)
as well as functions with a $1/x^{1.5}$
singularity (for instance the MRS[D$_-$] functions)
with and without shadowing.

\noindent
The hard process contributes above a $p_{\perp}$ threshold. The threshold is
set for less than three digits ISTRUF values to 3 GeV by default. This
option was used in the earlier DTUJET--92 version of the
program. An energy dependent cutoff\refmark{\BEKPR,\BEKPR} can
avoid a hard scattering cross section too large to be
treated in our simple eikonal approximation. To use this new
option the number 200 has to be added to the chosen ISTRUF value.
%
%
%
%
%
%
%

\bigskip
\noindent
Code word:  SEASU3

\line{Parameters used: \tt  SEASQ~=~WHAT(1) \hfil }

\noindent
This card concerns the fraction of strange quarks at the end of sea--chains.

\medskip
\line{\ \ \tt  SEASQ  \hfill  }

\vskip -0.25cm
\hangafter = 1
\hangindent = 2.1in
\parindent = 0pt
\leavevmode
\hbox to \hangindent
{\hskip 1.1in =~0.33~:\hfil}gives at the ends of
sea chains the same strangeness suppression as inside the chain decay.

\hangafter = 1
\hangindent = 2.1in
\parindent = 0pt
\leavevmode
\hbox to \hangindent
{\hskip 1.1in =~1.0~:\hfil}demands
equal fractions of u, d and s sea quarks at the ends of sea--chains.

\noindent
\line{The default is \tt 1. \hfil }

\vfill\eject
\noindent
Code word:  HADRONIZ

\line{Parameters used: \tt  IHADRZ~=~WHAT(1) \hfil }

\noindent
This card defines the fragmentation program used.

\medskip
\line{\ \ \tt  IHADRZ  \hfill  }

\vskip -0.5cm
\hangafter = 1
\hangindent = 2.1in
\parindent = 0pt
\leavevmode
\hbox to \hangindent
{\hskip 1.1in =~1~:\hfil}uses the BAMJET and DECAY programs.

\hangafter = 1
\hangindent = 2.1in
\parindent = 0pt
\leavevmode
\hbox to \hangindent
{\hskip 1.1in =~2~:\hfil}uses the program JETSET version 7.3 .

\noindent
\line{The default is {\tt 2}. \hfil }

\noindent
Most of our calculations used the BAMJET option.
The calculations in [{\BPER,\BEKPR}]  were done with the JETSET  option,
which allowed to include hard parton showers (see PSHOWER code word).

\noindent
The internal parameters of these programs are described in
the referenced papers\refmark\i  \refmark\sj ~. DTUJET--93 uses for
some JETSET parameters non-default values (see {\tt DTU93LUN.f} file),
DTUJET--93 also applies non-default values for some BAMJET parameters
(see {\tt DTU93TCB.f} file).

\medskip
\noindent
Code word:  POPCORN

\line{Parameters used: \tt  PDB=WHAT(1) \hfil }

\noindent
This card determines the role of diquarks in the fragmentation.

\medskip
\line{\ \ {\tt  PDB}  (with BAMJET) \hfill  }

\hangafter = 1
\hangindent = 2.1in
\parindent = 0pt
\leavevmode
\hbox to \hangindent
{\hskip 1.1in =~0~..~1~:\hfil}gives the fraction
of diquarks fragmenting in the
directly into baryons.

\medskip
\line{\ \ {\tt  PDB}  (with JETSET) \hfill  }
\hangafter = 1
\hangindent = 2.1in
\parindent = 0pt
\leavevmode
\hbox to \hangindent
{\hskip 1.1in =~0~:\hfil}The POPCORN mechanism is switched off.

\hangafter = 1
\hangindent = 2.1in
\parindent = 0pt
\leavevmode
\hbox to \hangindent
{\hskip 1.1in $>$~0~:\hfil}The POPCORN mechanism is switched on.

\noindent
\line{The default is {\tt 0.5}. \hfil }

\medskip
\noindent
Code word:  PSHOWER

\line{Parameters used: \tt  IPSHOW=WHAT(1) \hfil }

\noindent
This card determines whether hard partons initiate showers and is only
recognized in connection with JETSET fragmentation.
As the BAMJET option presently contains no
parton showering it does not reproduce
the $p_{\perp}$--distribution as well as the JETSET option.

\medskip
\line{\ \ {\tt  IPSHOW}  (with JETSET) \hfill  }

\hangafter = 1
\hangindent = 2.1in
\parindent = 0pt
\leavevmode
\hbox to \hangindent
{\hskip 1.1in =~0~:\hfil}Generation of hard parton showers suppessed.

\hangafter = 1
\hangindent = 2.1in
\parindent = 0pt
\leavevmode
\hbox to \hangindent
{\hskip 1.1in =~1~:\hfil}Hard parton showers are  included.

\noindent
\line{The default is {\tt 1}. \hfil }


\vfill\eject
\noindent
Code word:  START

\line{Parameters used: \tt  NEVNT~=~WHAT(1), PTLAR~=~WHAT(5) \hfil }

\noindent
Starts the sampling of hadronized events and the calculation
of the standard histogram output.

\hangafter = 1
\hangindent = 2.1in
\parindent = 0pt
\leavevmode
\hbox to \hangindent
{\hskip 0.2in \tt  NEVNT~\hfil}Number of events sampled. \hfil

\hangafter = 1
\hangindent = 2.1in
\parindent = 0pt
\leavevmode
\hbox to \hangindent
{\hskip 0.2in \tt  PTLAR~\hfil}Cutoff parameter
to sample only selected events. \hfill

\hangafter = 1
\hangindent = 2.1in
\parindent = 0pt
\leavevmode
\hbox to \hangindent
{\hskip 1.1in =~0.0:\hfil}DTUJET samples without constraint on sampled events.

\hangafter = 1
\hangindent = 2.1in
\parindent = 0pt
\leavevmode
\hbox to \hangindent
{\hskip 1.1in $>$~2.2:\hfil}DTUJET samples only events with at least one
jet (minijet) with $p_{\perp} \geq $ {\tt PTLAR}, rejecting all other events.

\hangafter = 1
\hangindent = 2.1in
\parindent = 0pt
\leavevmode
\hbox to \hangindent
{\hskip 1.1in $<$~-0.1:\hfil}DTUJET samples
only events without hard jets (minijets).

\smallskip
\noindent
{Defaults are not effective, as this card is necessary to run the program.
The code word START should be used only once per run.
\hfil }

\bigskip

\noindent
Code word: EVENTAPE

\line{Parameter used:  none\hfil }

\noindent
When this card is present, the events generated are
written on an output file {\tt EVENTS.DTU}.
This file is written in the following format:

\parindent 25pt
1.~card: {\tt ``This file contains events from DTUJET.''}

2.~card: The title of the input file is reproduced.

3.~card: {\tt NHAD} using {\tt FORMAT(I10)}

4. to $3+$ {\tt NHAD}
-th~card:~~~{\tt I, NR(I), PX(I), PY(I), PZ(I), E(I)\hfil}

\line{\hskip 2.0in using \tt FORMAT (2I4, 4E 16.8 )\hfil}

\vskip -0.3cm
\noindent
where

\hangafter = 1
\hangindent = 2in
\parindent = 0pt
\leavevmode
\hbox to \hangindent{\hskip 25pt {\tt NHAD }\hfil}is the
number of secondary particles in the event.

\hangafter = 1
\hangindent = 2in
\parindent = 0pt
\leavevmode
\hbox to \hangindent{\hskip 25pt {\tt I} and {\tt NR(I)}\hfil}are a counting
number and an integer label defining the kind of particle.

\hangafter = 1
\hangindent = 2in
\parindent = 0pt
\leavevmode
\hbox to \hangindent{\hskip 25pt {\tt PX(I), PY(I), PZ(I)}\hfil}are the
momentum
components of the secondary particles in the c.m.s.
of the collision in GeV/c.
The projectile is moving in positive $z$--direction.

\hangafter = 1
\hangindent = 2in
\parindent = 0pt
\leavevmode
\hbox to \hangindent{\hskip 25pt {\tt E(I)} \hfil}are the total energies
 of the secondary particles.
%

\medskip
\noindent

The codes for stable particles considered at the end are given in Table 1.

\vfill\eject
\noindent
{\bf Table 1:} List of the codes for stable particles used in
BAMJET and DTUJET93.
\medskip
\begintable
$  1 = p  $                |   $ 14 = \pi^- $             | $
  98 = \Xi^-                 $ |  $  121= D^-_s                 $ \cr
$  2 = \bar p $            |   $ 15 = K^+ $               | $
  99 = \bar\Sigma^-          $ |  $  137= \Lambda^+_c           $ \cr
$  3 = e^+ $               |   $ 16 = K^- $               | $
  100= \bar\Sigma^0          $ |  $  149= \bar \Lambda^+_c      $ \cr
$  4 = e^- $               |   $ 17 = \Lambda $           | $
  101=\bar\Sigma^+           $ |  $  138= \Xi^+_c               $ \cr
$  5 = \nu $               |   $ 18 = \bar \Lambda $      | $
  102= \bar\Xi^0             $ |  $  139= \Xi^0_c               $ \cr
$  6 = \bar \nu $          |   $ 19 = K^0_S $             | $
  103= \bar\Xi^+             $ |  $  150= \bar\Xi^+_c           $ \cr
$  7 = \gamma $            |   $ 20 = \Sigma^- $          | $
  109= \Omega^-              $ |  $  151= \bar\Xi^0_c           $ \cr
$  8 = n $                 |   $ 21 = \Sigma^+ $          | $
  115=\bar\Omega^-           $ |  $  140= \Sigma^{++}_c         $ \cr
$  9 = \bar n $            |   $ 22 = \Sigma^0 $          | $
  116= D^0                   $ |  $  141= \Sigma^+_c            $ \cr
$ 10 = \mu^+ $             |   $ 23 = \pi^0 $             | $
  117= D^+                   $ |  $  142= \Sigma^0_c            $ \cr
$ 11 = \mu^- $             |   $ 24 = K^0 $               | $
  118= D^-                   $ |  $  152= \bar\Sigma^{++}_c     $ \cr
$ 12 = K^0_L $             |   $ 25 = \bar K^0  $         | $
  119= \bar D^0              $ |  $  153= \bar\Sigma^+_c        $ \cr
$ 13 = \pi^+ $             |   $  97 = \Xi^0 $            | $
  120= D^+_s                 $ |  $  154= \bar\Sigma^0_c        $\endtable

For a list of particles known to DTUJET see the PARTICLE
input card.
A integer array {\tt NBOOK(NR)} to convert our internal number {\tt NR}
below {\tt NR}=25 to the particle data booklet number
is provided in {\tt BLOCK DATA BOOKLT} at the end of {\tt dtu93mai}.
This conversion is done for all the stable
particles given above by the {\tt INTEGER FUNCTION MPDGHA(NR)}
included in the {\tt dtu93lun.f} file.

\medskip
\noindent
Code word:  STOP

\line{Parameter used: none\hfil }

\noindent
Stops the running of the code.

\vfill\eject
\bigskip
\noindent
Code word:  PARTICLE

\line{Parameter used:  none\hfil }

\noindent
This card demands the printing of a table of all particles
defined in DTUJET where the particle properties and
in the case of unstable particles all decay channels defined
in the code are listed.  The particles defined in DTUJET are the ones defined
in the BAMJET and DECAY hadronization codes.

At present the normal output events of DTUJET contain
particles which do not decay through hadronic interactions (most of them
with number labels below {\tt NR}~=~26).
This selection can be changed by modifying the
defaults in the {\tt SUBROUTINE DECAY} code of the {\tt dtu93tcb.f} file
(in the case of BAMJET fragmentation)
and in the {\tt dtu93lun.f} file (in the case of JETSET fragmentation).

\bigskip
\noindent
Code word:  XSECTION

\line{Parameter used:  none\hfil }

\noindent
This card demands a test run for the calculation of cross sections as
function of energy.
The test run yields

\medskip
\line{\hskip 0.4in \ the total cross section,                           \hfil}
\line{\hskip 0.4in \ the elastic cross section,                          \hfil}

\line{\hskip 0.4in \ the inelastic cross section,                        \hfil}

\line{\hskip 0.4in \ the hard inelastic cross section and                \hfil}

\line{\hskip 0.4in \ the high and low mass \hfil}
\line{\hskip 0.4in \ \ \ \ single and double diffractive cross sections.  \hfil
}

It also provides scatter plots of the sampling of multi-Pomeron
events at some typical energies.

\bigskip
\noindent
Other code words as LHARD, SIGMAPOM, PARTEV etc. are available for
special tasks but not relevant for the general user. The interested
user is referred to the comments in the source code (in {\tt dtu93mai.f}).

\vfill\eject

\section{  Common Blocks of Interest for the User}

It is possible to run the event generator within a user supplied
program package. For this purpose the user needs to know
how the event is stored in COMMON blocks.

\bigskip
The COMMON  block /USER/

\medskip
\line{\tt \hskip 0.4in \ \
CHARACTER*8 PROJTY,TARGTY                         \hfil}

\line{\tt \hskip 0.4in \ \
CHARACTER*80 TITLE                                \hfil}

\line{\tt \hskip 0.4in \ \
COMMON /USER/TITLE,PROJTY,TARGTY,CMENER,ISTRUF   \hfil}

\line{\tt \hskip 0.4in \& \           ,
ISINGD,IDUBLD,SDFRAC,PTLAR            \hfil}

contains the parameters, expected to be modified by input cards.
The meaning of the possible values of
these parameters are described in section 3 .

The COMMON block /PARTCL/

\noindent
\line{\tt \hskip 0.4in \ \  PARAMETER (MXNUPA=2500) \hfil}
\line{\tt \hskip 0.4in \ \  COMMON/PARTCL/ \hfil}
\line{\tt \hskip 0.4in
\& \ PARTPX(MXNUPA),PARTPY(MXNUPA), \hfil}
\line{\tt \hskip 0.4in
\& \ PARTPZ(MXNUPA),PARTP0(MXNUPA), \hfil}
\line{\tt \hskip 0.4in \& \ NPARTY(MXNUPA),NPART \hfil}

contains the final particles of the last generated event. The first NPART's
positions in the arrays

\medskip
\line{\tt \hskip 0.4in \ \ PARTPX,PARTPY,PARTPZ,PARTP0,NPARTY \hfil}

are transverse momenta, longitudinal momentum, energy and
the type of the particles where

\medskip
\line{\tt \hskip 0.4in \ \ NPART \hfil}

is number of particles in the event.
%
%
%
%
%
%

For how to obtain a printed list of our particle and resonance numbering
(i.e. the meaning of {\tt NPARTY})
see the input card {\tt PARTICLE}. For a short list and for ways to convert to
the particle data booklet numbers we refer to the discussion of the
input card {\tt EVENTAPE}.
We did not use the particle data booklet numbers, as their range
is too large to use them in better way than alphanumeric names.

\vfill\eject
\section{ Short Description of Important Subroutines of DTUJET}

\line{\it 3.5.1 ~Subroutine calls for generating  events \hfil}

The main task of that program ist to generate events. This involves the
following five steps:

\item{1.}
The program begins with a call to the subroutine DTPREP, which sets
initial values and reads and implements the input cards.

\item{2.}
After encountering the START input card
the sampling of multi-Pomeron (hard and soft Pomerons) distributions is
prepared by a call to PRBLM2. It calculates the probability
distribution of hard and soft Pomerons and stores it in a suitable way. A call
to
SAMPLM then generates one basic scattering event according to this
distribution.
Various calls to other subroutines,  which prepare other aspects of the
sampling follow. After these tasks the subroutine
DTPREP is eventually left to the main program.

\item{3.}
In the main program the loop over events follows.
Within this event loop
a subroutine DTCOLL is called, which generates one complete hadronic event
(see below). After the end of the loop and a short output
the program returns to the beginning and continues
until a STOP card is encountered.

\item{4.}
Within DTCOLL a call to XPTFL (NHARD, NSEA, NVAL)  generates one
parton level event with NHARD hard Pomerons and NSEA soft sea
Pomerons and one (=NVAL) soft valence Pomeron. It determines the
energy-momentum distribution of the partons.

\item{5.}
The hadronization of this parton level event is done by a call from
DTCOLL to the routine HADRAV (NHAD, NSEA, NVAL, NHARD).
The event generated contains NHAD particles in its final state.
Depending on the parameter IHADRZ set with the HADRONIZ parameter card
the BAMJET and DECAY routines or the JETSET routines are used
for fragmentation.
%

\bigskip
\line{\it 3.5.2 ~Calculation of total and inelastic cross sections and\hfil}
\line{\hskip 2.6pc \it hard and soft Pomeron distributions
via the unitarization of\hfil}
\line{\hskip 2.6pc \it soft and hard input cross sections\hfil}

\parindent=25pt

The code word  XSECTION demands a major executable task
(used essentially for test purposes) which calculates
absorbed cross sections in the DTU model.

When demanded
by the XSECTION input card the subroutine DTPREP calls
the subroutine POMDI. POMDI calls the routine SIGMAS
to calculate the total and inelastic
cross sections according to the model as function of energy.
The input hard, soft, and triple Pomeron
cross sections of a given energy are defined by a call to the subroutine
SIGSHD.
The output cross sections are presented as alphanumeric plots.

\vfill\eject
The sampling of multi-Pomeron (hard and soft Pomeron) distributions is
prepared by a call to PRBLM2, which calculates the probability
distribution for $n_h$ hard and $n_s$ soft Pomerons.  A call to
SAMPLM generates one event from this distribution.  POMDI generates
scatter plots in $n_h$ and $n_s$ for 10000 sampled events
at some collision energies.

%
%

\section{Exemplary Input and Command Files}

In Fig.~1 we reproduce an input file which allows to run DTUJET--93.
This file, where we include some possible input cards even when they
do not change the default settings, will lead to the calculation of
5000 events at $\sqrt s$~= 1.8 TeV generating standard histograms.

In Fig.~2a we list an exemplary command file used for compiling
DTUJET--93 files
under DEC--ULTRIX, the same is given in Fig.~2b and c for HP--UX
and IBM AIX respectively. Please note,
that the code, not written explicitly in DOUBLE PRECISION is transformed
to DOUBLE PRECISION using the compiler option for the redefinition of the
constants and D-commented ("debug"--) cards in the files.

\section{Description of the Default Short Output}

In the papers [{\Pertermann,\Engel,\Roesler,\BPER,\RBEP}]
many plots are shown which originate
from the DTUJET--93 code; therefore we do
not need to give explicit examples for
the output of DTUJET.

After repeating the input cards the minimal output contains
a table of all calculated cross sections. A table containing the
average multiplicity and
the average energy separated for specific particle types
follows. Simple alpha numeric plots then
draw the rapidity, energy fraction and multiplicity distribution.

\vfill
\centerline{\bf Acknowledgements}

\parindent=25pt

We would like to thank our collaborators
A.~Capella, J.~Kwiecinski, K.~Hahn, M.~Maire
 and J.~Tran Thanh Van for discussions and a collaboration, which led to
 the underlying model.
Two of the authors (R.E. and D.P) acknowledge support by the Deutsche
Forschungsgemeinschaft under contract Ra 559/3-1 and
under contract  Pe 540/1-1.

\vskip 2cm
\refout

\vfill\eject

\bigskip \noindent
\line{\tt \hskip 0.2in
TITLE~~~~~~~~~~~~~~~~~~~~~~~~~~~~~~~~~~~~~~~~~~~~~~~~~~~~~~~~~~~~~~~~~~\hfill}
\line{\tt \hskip 0.2in
Antiproton-Proton with diffraction, STRUCFUN 215 and JETSET
fragmentation.\hfill}
\line{\tt \hskip 0.2in
CMENERGY~~~~~~1800.~~~~~~~~~~~~~~~~~~~~~~~~~~~~~~~~~~~~~~~~~~~~~~~~~~~~\hfill}
\line{\tt \hskip 0.2in
PROJPAR~~~~~~~~~~~~~~~~~~~~~~~~~~~~~~~~~~~~~~~~~~~~~~~~~~~~~~~~~APROTON\hfill}
\line{\tt \hskip 0.2in
TARPAR~~~~~~~~~~~~~~~~~~~~~~~~~~~~~~~~~~~~~~~~~~~~~~~~~~~~~~~~~~PROTON~\hfill}
\line{\tt \hskip 0.2in
SINGDIFF~~~~~~~~~1.~~~~~~~0.~~~~~1.00~~~~~~~~~~~~~~~~~~~~~~~~~~~~~~~~~~\hfill}
\line{\tt \hskip 0.2in
START~~~~~~~~~5000.~~~~~~~0.~~~~~~~0.~~~~~~~0.~~~~~~~2.0~~~~~~~~~~~~~~~\hfill}
\line{\tt \hskip 0.2in STOP\hfill }

\bigskip \noindent
\line{{\bf Fig~1:~~}DTUJET--93 sample input file.\hfill }

\bigskip \noindent

\bigskip \noindent
\line{\tt \hskip 0.2in
f77~~-c~-diag~-fpe2~-o0~-K~-d$\_$lines~-r8~-C~-V~<file name>.f\hfill}

\bigskip \noindent
\line{{\bf Fig~2a:~~}Command file for
compiling DTUJET--93 Fortran files under DEC--ULTRIX.\hfill }

\bigskip \noindent

\bigskip \noindent
\line{\tt \hskip 0.2in
f77~~-c~-a~+e~~-K~-D~-R8~~+T~+R~-C~-V~<file name>.f\hfill}

\bigskip \noindent
\line{{\bf Fig~2b:~~}Command file for compiling
DTUJET--93 Fortran files under HP--UX.\hfill }

\bigskip \noindent

\bigskip \noindent
\line{\tt \hskip 0.2in
xlf~~-C~-c~-qsource -qdpc -D~<file name>.f \hfill}

\bigskip \noindent
\line{{\bf Fig~2c:~~}Command file
for compiling DTUJET--93 Fortran files under IBM-AIX.\hfill }

\nobreak
\vfill\eject\end

\bye